\documentclass[nofootinbib,prd]{revtex4}%
\usepackage[latin9]{inputenc}
\usepackage{float}
\usepackage{amsmath}
\usepackage{amssymb}
\usepackage{graphicx}
\usepackage{esint}
\usepackage{amsfonts}%
\makeatletter
\@ifundefined{textcolor}{}
{
\definecolor{BLACK}{gray}{0}
\definecolor{WHITE}{gray}{1}
\definecolor{RED}{rgb}{1,0,0}
\definecolor{GREEN}{rgb}{0,1,0}
\definecolor{BLUE}{rgb}{0,0,1}
\definecolor{CYAN}{cmyk}{1,0,0,0}
\definecolor{MAGENTA}{cmyk}{0,1,0,0}
\definecolor{YELLOW}{cmyk}{0,0,1,0}
}
\makeatother
\begin{document}
\title{Rainbow's Stars}
\author{Remo Garattini}
\email{Remo.Garattini@unibg.it}
\affiliation{Universit\`{a} degli Studi di Bergamo, Department of Engineering and Applied
Sciences, Viale Marconi 5, 24044 Dalmine (Bergamo) Italy.}
\affiliation{I.N.F.N. - sezione di Milano, Milan, Italy.}
\author{Gianluca Mandanici}
\email{Gianluca.Mandanici@unibg.it}
\affiliation{Universit\`{a} degli Studi di Bergamo, Department of Engineering and Applied
Sciences, Viale Marconi 5, 24044 Dalmine (Bergamo) Italy.}

\begin{abstract}
In recent years, a growing interest on the equilibrium of compact astrophysical objects like white dwarf and neutron stars has been manifested. 
In particular, various modifications due to Planck scale energy effects have been considered. In this paper we analyze the modification induced by
 Gravity's Rainbow on the equilibrium configurations described by the Tolman-Oppenheimer-Volkoff (TOV) equation. 
Our purpose is to explore the possibility that the Rainbow Planck-scale deformation of space-time could support the existence of different compact stars.

\end{abstract}
\maketitle

\section{Introduction}

Compact stars, exotic stars, wormholes and black holes are astrophysical
objects described by the Einstein's Field equations. For a perfect fluid and
in case of spherical symmetry, these objects obey the
Tolman-Oppenheimer-Volkoff (TOV) equation (in c.g.s. units) \cite{Tolman,OV}%
\begin{equation}
\frac{dp_{r}\left(  r\right)  }{dr}=-\left(  \rho\left(  r\right)
+\frac{p_{r}\left(  r\right)  }{c^{2}}\right)  \frac{4\pi Gr^{3}p_{r}\left(
r\right)  /c^{2}+Gm(r)}{r^{2}\left[  1-2Gm(r)/rc^{2}\right]  }+\frac{2}%
{r}\left(  p_{t}\left(  r\right)  -p_{r}\left(  r\right)  \right)  \label{TOV}%
\end{equation}
and%
\begin{equation}
\frac{dm}{dr}=4\pi\rho\left(  r\right)  r^{2},
\end{equation}
where $c$ is the speed of light, $G$ is the gravitational constant,
$\rho\left(  r\right)  $ is the macroscopic energy density measured in proper
coordinates and $p_{r}\left(  r\right)  $ and $p_{t}\left(  r\right)  $ are
the radial pressure and the transverse pressure, respectively. It is clear
that the knowledge of $\rho\left(  r\right)  $ allows to understand the
astrophysical structure under examination. If we fix our attention on compact
stars, ordinary General Relativity offers two kind of exact solutions for the
isotropic TOV equation:

\begin{description}
\item[a)] the constant energy density solution,

\item[b)] the Misner-Zapolsky energy density solution\cite{MisnerZapolsky}.
\end{description}

Of course, a) and b) can be combined to give a new profile which has been
considered by Dev and Gleiser\cite{DevGleiser}. \noindent The case b) is
satisfied with the help of an equation of state of the form $p_{r}=\omega\rho$
with $\omega=1/3$%
\begin{equation}
p_{r}=\omega\rho\left(  r\right)  =\omega\frac{3c^{2}}{56\pi Gr^{2}}%
=\frac{c^{2}}{56\pi Gr^{2}}\label{MZ}%
\end{equation}
and%
\begin{equation}
m(r)=\frac{3c^{2}r}{14G}.
\end{equation}
Other different solutions can be found introducing
anisotropy\cite{DevGleiser,KRHR} and/or polytropic transformations\cite{FM} or
other forms of modification of gravity like $f\left(  R\right)  $
gravity\cite{CSfR, CSfRT} and Generalized Uncertainty Principle
(GUP)\cite{GUP}. The GUP\ distortion is only one of the different examples
involving Planckian or Trans-Planckian modifications due to quantum
gravitational effects coming into play. Indeed, a number of recent studies
have already focused on the effects of Planck-scale physics on the equilibrium
configuration of compact astrophysical objects (see e.g.
\cite{Camacho:2006qg,Gregg:2008jb,AmelinoCamelia:2009tv,AmelinoCamelia:2010rm,Wang:2010ct,Carmona:2013fwa,Rovelli:2014cta,Barrau:2014hda,Ali:2014xqa}%
). Usually Planck scale physics is considered to affect equilibrium
configuration via the modification of the energy-momentum dispersion relation
that implies deformed equation of state (EoS) for the fluid composing the
star. This is for example the approach followed in Refs.
\cite{Camacho:2006qg,Gregg:2008jb,AmelinoCamelia:2009tv,AmelinoCamelia:2010rm}%
. However there are Planck scale scenarios in which the deformation occurs by
means of the metric deformation as well. This is the case of the so called
Gravity's Rainbow\cite{Magueijo:2001cr,Magueijo:2002xx}. Gravity's Rainbow is
a distortion of space-time induced by two arbitrary functions, $g_{1}\left(
E/E_{\mathrm{Pl}}\right)  $ and $g_{2}\left(  E/E_{\mathrm{Pl}}\right)  $,
which have the following property\footnote{Applications and implications of
Gravity's Rainbow in Astrophysics and cosmology can be found
in\cite{Garattini:2009nq,Garattini:2011fs, RGGM1,RGGM, Garattini:2012ec,
Garattini:2013yha}.}
\begin{equation}
\lim_{E/E_{\mathrm{Pl}}\rightarrow0}g_{1}\left(  E/E_{\mathrm{Pl}}\right)
=1\qquad\text{and}\qquad\lim_{E/E_{\mathrm{Pl}}\rightarrow0}g_{2}\left(
E/E_{\mathrm{Pl}}\right)  =1.\label{lim}%
\end{equation}
It has been introduced for the first time by Magueijo and
Smolin\cite{Magueijo:2001cr, Magueijo:2002xx}, who proposed that the
energy-momentum tensor and the Einstein's Field Equations were modified with
the introduction of a one parameter family of equations
\begin{equation}
G_{\mu\nu}\left(  E/E_{\mathrm{Pl}}\right)  =8\pi G\left(  E/E_{\mathrm{Pl}%
}\right)  T_{\mu\nu}\left(  E/E_{\mathrm{Pl}}\right)  +g_{\mu\nu}%
\Lambda\left(  E/E_{\mathrm{Pl}}\right)  ,\label{Gmn}%
\end{equation}
where $G\left(  E/E_{\mathrm{Pl}}\right)  $ is an energy dependent Newton's
constant and $\Lambda\left(  E/E_{\mathrm{Pl}}\right)  $ is an energy
dependent cosmological constant, defined so that $G\left(  0\right)  $ is the
low-energy Newton's constant and $\Lambda\left(  0\right)  $ is the low-energy
cosmological constant. It is clear that the modified Einstein's Field
equations $\left(  \ref{Gmn}\right)  $ give rise to a class of solutions which
are dependent on $g_{1}\left(  E/E_{\mathrm{Pl}}\right)  $ and $g_{2}\left(
E/E_{\mathrm{Pl}}\right)  $. For instance, the \textit{rainbow} version of the
Schwarzschild line element is%
\begin{equation}
ds^{2}=-\left(  1-\frac{2MG\left(  0\right)  }{r}\right)  \frac{d\tilde{t}%
^{2}}{g_{1}^{2}\left(  E/E_{\mathrm{Pl}}\right)  }+\frac{d\tilde{r}^{2}%
}{\left(  1-\frac{2MG\left(  0\right)  }{r}\right)  g_{2}^{2}\left(
E/E_{\mathrm{Pl}}\right)  }+\frac{\tilde{r}^{2}}{g_{2}^{2}\left(
E/E_{\mathrm{Pl}}\right)  }\left(  d\theta^{2}+\sin^{2}\theta d\phi
^{2}\right)  .\label{line}%
\end{equation}
As shown in Ref.\cite{RGGM,RGGM1}, one of the effects of the functions
$g_{1}\left(  E/E_{P}\right)  $ and $g_{2}\left(  E/E_{P}\right)  $ is to keep
under control UV\ divergences allowing therefore the computation of quantum
corrections to classical quantities, at least to one loop. As a result, the
computation of Zero Point Energy (ZPE) in \textit{Gravity's Rainbow} is well
defined for appropriate choices of $g_{1}\left(  E/E_{P}\right)  $ and
$g_{2}\left(  E/E_{P}\right)  $. In this paper, we would like to consider the
effect of \textit{Gravity's Rainbow} on the TOV equations to explore the
possibility of finding new forms of compact stars. The paper is organized as
follows. In section \ref{p2} we consider the TOV modified by Gravity's
Rainbow, in section \ref{p3} we examine the constant energy density case and
its consequence on the redshift factor, in section \ref{p4} we examine the
variable energy density case and its consequence on the redshift factor
including the Dev-Gleiser case. We summarize and conclude in section \ref{p5}.

\section{TOV Equation in Gravity's Rainbow}

\label{p2}To see how Gravity's Rainbow affects the TOV equations, we need to
define the following line element%
\begin{equation}
ds^{2}=-\frac{e^{2\Phi(r)}}{g_{1}^{2}(E/E_{\mathrm{Pl}})}c^{2}dt^{2}%
+\frac{dr^{2}}{g_{2}^{2}(E/E_{\mathrm{Pl}})\left(  1-\frac{2Gm(r)}{rc^{2}%
}\right)  }+\frac{r^{2}}{g_{2}^{2}(E/E_{\mathrm{Pl}})}\left(  d\theta^{2}%
+\sin^{2}\theta d\phi^{2}\right)  . \label{eq:metrica-1}%
\end{equation}
From Appendix \ref{App1}, we can see that only $G_{00}$ modifies:%
\begin{equation}
G_{00}=2G\frac{e^{2\Phi(r)}}{r^{2}}\frac{g_{2}^{2}(E/E_{\mathrm{Pl}})}%
{g_{1}^{2}(E/E_{\mathrm{Pl}})}m^{\prime}(r).
\end{equation}
For the energy-momentum stress tensor describing a perfect-fluid, we assume
the following form%
\begin{equation}
T_{\mu\nu}=\left(  \rho c^{2}+p_{t}\right)  u_{\mu}u_{\nu}+p_{t}g_{\mu\nu
}+\left(  p_{r}-p_{t}\right)  n_{\mu}n_{\nu},
\end{equation}
where $u^{\mu}$ is the four-velocity normalized in such a way that $g_{\mu\nu
}u^{\mu}u^{\nu}=-1$, $n_{\mu}$ is the unit spacelike vector in the radial
direction, i.e. $g_{\mu\nu}n^{\mu}n^{\nu}=1$ with $n^{\mu}=\sqrt{1-2Gm\left(
r\right)  /rc^{2}}\delta_{r}^{\mu}$. $\rho\left(  r\right)  $ is the energy
density, $p\left(  r\right)  $ is the radial pressure measured in the
direction of $n^{\mu}$, and $p_{t}\left(  r\right)  $ is the transverse
pressure measured in the orthogonal direction to $n^{\mu}$. From the results
of the Appendix \ref{App1}, we can see that the equilibrium equation%
\begin{equation}
\frac{dp}{dr}+\left(  \rho c^{2}+p\right)  \Phi^{\prime}(r)=0 \label{EqEq}%
\end{equation}
must hold also in Gravity's Rainbow. From this equation follows that%
\begin{equation}
\frac{dp_{r}}{dr}=-\left(  \rho+\frac{p_{r}}{c^{2}}\right)  \frac{\kappa
r^{3}p_{r}/c^{4}g_{2}^{2}(E/E_{\mathrm{Pl}})+2Gm(r)/c^{2}}{2r^{2}\left[
1-2Gm(r)/rc^{2}\right]  }+\frac{2}{r}\left(  p_{t}-p_{r}\right)
\label{TOVGRw}%
\end{equation}
and%
\begin{equation}
\frac{dm}{dr}=\frac{4\pi\rho(r)r^{2}}{g_{2}^{2}(E/E_{\mathrm{Pl}})},
\label{m'(r)}%
\end{equation}
where $\rho$ is the mass density. Eq.$\left(  \ref{TOVGRw}\right)  $ is the
anisotropic TOV equation modified by Gravity's Rainbow. As a first
simplification, we will assume that the star is isotropic. Then, we will
consider the constant energy density case I) and the Misner-Zapolsky energy
density case II). We begin to consider the case I).

\section{Isotropic pressure and the constant energy density case}

\label{p3}With the assumption of an isotropic star, the pressure in
Eq.$\left(  \ref{TOVGRw}\right)  $ becomes%
\begin{equation}
\frac{dp_{r}}{dr}=-\left(  \rho+\frac{p_{r}(r)}{c^{2}}\right)  \frac{4\pi
Gr^{3}p_{r}(r)/c^{2}g_{2}^{2}(E/E_{\mathrm{Pl}})+Gm(r)}{r^{2}\left[
1-2Gm(r)/rc^{2}\right]  }. \label{RTOV2}%
\end{equation}
The constant energy density assumption allows an easy solution of Eq.$\left(
\ref{m'(r)}\right)  $. Indeed, one gets%
\begin{equation}
m(r)=\frac{4\pi\rho}{3g_{2}^{2}(E/E_{\mathrm{Pl}})}r^{3}, \label{eq:mg2}%
\end{equation}
where we have used the boundary conditions $m(0)=0$. Nevertheless,
Eqs.$\left(  \ref{RTOV2}\right)  $ and $\left(  \ref{eq:mg2}\right)  $ are
referred to the whole star included the external boundary $R$. To account for
different scenarios we discuss two fundamental cases:

\begin{description}
\item[a)] The star is divided in two regions: the inner region or the core,
where Gravity's Rainbow is relevant, and the outer region, where Gravity's
Rainbow is negligible.

\item[b)] The whole star is modified by Gravity's Rainbow.
\end{description}

\subsubsection{ Case a)}

In this case, the star is divided in two parts: the external part of the star
without Gravity's Rainbow and the core with Gravity's Rainbow. Basically, we
can write%
\begin{equation}
dm=\left\{
\begin{array}
[c]{cc}%
4\pi\rho r^{2}dr/g_{2}^{2}(E/E_{\mathrm{Pl}}) & \bar{r}\geq r>0\\
4\pi\rho r^{2}dr & R>r>\bar{r}%
\end{array}
\right.  . \label{dm}%
\end{equation}
The transition between the distorted and the undistorted mass is represented
by introducing an intermediate radius $\bar{r}$, assuming that%
\begin{equation}
R\ggg\bar{r}>l_{\mathrm{Pl}}. \label{radius}%
\end{equation}
In this first approach, the transition between the distorted and the
undistorted mass is very sharp, but we cannot exclude the possibility of
describing a smoothed variation between the external part of the star and the
core in a next future. After an integration, we can write%
\begin{equation}
m\left(  r\right)  =\left\{
\begin{array}
[c]{cc}%
m_{1}\left(  r\right)  =\frac{4\pi\rho r^{3}}{3g_{2}^{2}(E/E_{\mathrm{Pl}}%
)}=Mr^{3}/\left(  \tilde{R}^{3}g_{2}^{2}(E/E_{\mathrm{Pl}})\right)  & \bar
{r}\geq r>0\\
m_{2}\left(  r\right)  =\frac{4\pi\rho}{3}\left(  r^{3}+\bar{r}^{3}A\left(
E/E_{\mathrm{Pl}}\right)  \right)  =M\left(  r^{3}+\bar{r}^{3}A\left(
E/E_{\mathrm{Pl}}\right)  \right)  /\tilde{R}^{3} & R>r>\bar{r}%
\end{array}
\right.  . \label{M}%
\end{equation}
In Eq.$\left(  \ref{M}\right)  $ we have used the total mass density%
\begin{equation}
\rho=M\left[  \frac{4\pi}{3}\left(  R^{3}+A\left(  E/E_{\mathrm{Pl}}\right)
\bar{r}^{3}\right)  \right]  ^{-1}=M/\tilde{V} \label{rhoM}%
\end{equation}
and we have defined%
\begin{equation}
\tilde{V}=\frac{4\pi}{3}\left(  R^{3}+A\left(  E/E_{\mathrm{Pl}}\right)
\bar{r}^{3}\right)  =\frac{4\pi}{3}\tilde{R}^{3}, \label{Vol}%
\end{equation}
with%
\begin{equation}
\tilde{R}^{3}=R^{3}+A\left(  E/E_{\mathrm{Pl}}\right)  \bar{r}^{3}
\label{Rtilde}%
\end{equation}
and%
\begin{equation}
A\left(  E/E_{\mathrm{Pl}}\right)  =\left(  g_{2}^{2}(E/E_{\mathrm{Pl}}%
)^{-1}-1\right)  .
\end{equation}
We indicate with $\rho_{0}$, the mass density $\left(  \ref{rhoM}\right)  $
with $g_{2}(E/E_{\mathrm{Pl}})=1$. Note that the volume distorted by Gravity's
Rainbow, for a sphere of radius $R$, is%
\begin{equation}
V=\int d^{3}x\sqrt{g}=\frac{4\pi}{3g_{2}^{3}(E/E_{\mathrm{Pl}})}\int_{0}%
^{R}\frac{r^{2}dr}{\sqrt{1-2m\left(  r\right)  /r}}.
\end{equation}
Therefore the mass density in $\left(  \ref{rhoM}\right)  $ does not coincide
with the ratio $M/V$. To calculate the pressure, we divide the radius of the
star into two sectors exactly like in Eq. $\left(  \ref{dm}\right)  $. We
begin to consider the range $R\geq r\geq\bar{r}$. This is the sector where the
TOV equation is undeformed. From Eq.$\left(  \ref{RTOV2}\right)  $, with
$g_{2}(E/E_{\mathrm{Pl}})=1$, one gets%
\begin{equation}
p_{r}(r)=\rho_{0}c^{2}\frac{\left(  \sqrt{3c^{2}-\kappa\rho_{0}r^{2}}%
-\sqrt{3c^{2}-\kappa\rho_{0}R^{2}}\right)  }{\left(  3\sqrt{3c^{2}-\kappa
\rho_{0}R^{2}}-\sqrt{3c^{2}-\kappa\rho_{0}r^{2}}\right)  }, \label{p(r)Out}%
\end{equation}
where $\kappa=8\pi G$ and where we have used the boundary condition $p\left(
R\right)  =0$. It is immediate to recognize that in this region of the star,
to avoid a singularity in the denominator, we have to impose%
\begin{equation}
R<\sqrt{\frac{c^{2}}{3\pi G\rho_{0}}}. \label{ineq}%
\end{equation}
When we use the relationship $\left(  \ref{eq:mg2}\right)  $ with
$g_{2}(E/E_{\mathrm{Pl}})=1$, then we recover the Buchdahl-Bondi bound
\cite{BuchdahlPR59,Buchdahl1966ApJ,Bondi,Islam1969MNRAS}
\begin{equation}
M<\frac{4}{9}\frac{c^{2}}{G}R. \label{BB}%
\end{equation}
However, because of the distortion introduced by Gravity's Rainbow in
Eq.$\left(  \ref{eq:mg2}\right)  $ and in the mass density $\rho$ $\left(
\ref{rhoM}\right)  $, the inequality $\left(  \ref{BB}\right)  $ becomes%
\begin{equation}
M<\frac{4c^{2}}{9GR^{2}}\left(  R^{3}+A\left(  E/E_{\mathrm{Pl}}\right)
\bar{r}^{3}\right)  \label{BBm}%
\end{equation}
and the Buchdahl-Bondi bound is modified. It is useful to consider the limit
in which $E/E_{\mathrm{Pl}}\rightarrow0$. In this limit, we find that
Eq.$\left(  \ref{BBm}\right)  $ reduces to%
\begin{equation}
M<\frac{4c^{2}}{9GR^{2}}\left(  R^{3}+\left(  \left(  1+h(E/E_{\mathrm{Pl}%
})\right)  ^{-1}-1\right)  \bar{r}^{3}\right)  \simeq\frac{4c^{2}R}%
{9G}-h(E/E_{\mathrm{Pl}})\frac{4c^{2}\bar{r}^{3}}{9GR^{2}},
\end{equation}
where $h\left(  E/E_{\mathrm{Pl}}\right)  \rightarrow0$, when
$E/E_{\mathrm{Pl}}\rightarrow0$. Note that $h\left(  E/E_{\mathrm{Pl}}\right)
\gtrless0$ depending on the form of the rainbow's function. To complete the
analysis, we have to examine the core of the star $\bar{r}\geq r\geq0$ where
Gravity's Rainbow is switched on, leading to the following TOV equation%
\begin{equation}
\frac{dp_{r}}{dr}=-\frac{\kappa r\left(  \rho c^{2}+p(r)\right)  \left(
3p(r)+\rho c^{2}\right)  }{2c^{2}\left[  3c^{2}g_{2}^{2}(E/E_{\mathrm{Pl}%
})-\kappa\rho r^{2}\right]  }, \label{TOVin}%
\end{equation}
whose solution is%
\begin{equation}
p_{r}(r)=\rho c^{2}\frac{CB\left(  r,E\right)  -1}{3-CB\left(  r,E\right)  },
\label{p(r)<}%
\end{equation}
where $A$ is a constant to be determined by an appropriate choice of the
boundary conditions and where%
\begin{equation}
B\left(  r,E\right)  =\sqrt{3c^{2}g_{2}^{2}(E/E_{\mathrm{Pl}})-\kappa\rho
r^{2}}.
\end{equation}
Since $p_{r}(r)$ must be continuous, we have to impose%
\begin{equation}
\lim_{r-\bar{r}_{-}}p_{r}(r)=\lim_{r-\bar{r}_{+}}p_{r}(r)
\end{equation}
which implies%
\begin{equation}
\frac{CB\left(  \bar{r},E\right)  -1}{3-CB\left(  \bar{r},E\right)  }%
=\frac{\rho_{0}}{\rho}\frac{\left(  \sqrt{3c^{2}-\kappa\rho_{0}\bar{r}^{2}%
}-\sqrt{3c^{2}-\kappa\rho_{0}R^{2}}\right)  }{\left(  3\sqrt{3c^{2}-\kappa
\rho_{0}R^{2}}-\sqrt{3c^{2}-\kappa\rho_{0}\bar{r}^{2}}\right)  }=D\left(
\bar{r},R\right)  .
\end{equation}
Thus $C$ is no longer a constant but it has become a function of $\bar{r},R$
and $E$ and it is determined to find%
\begin{equation}
C\equiv C\left(  \bar{r},R,E\right)  =\frac{3D\left(  \bar{r},R\right)
+1}{B\left(  \bar{r},E\right)  \left(  1+D\left(  \bar{r},R\right)  \right)
}. \label{Match}%
\end{equation}
Plugging the value of $C\left(  \bar{r},R,E\right)  $ into $\left(
\ref{p(r)<}\right)  $, we obtain%
\begin{equation}
p_{r}(r)=\rho c^{2}\frac{\left(  3D\left(  \bar{r},R\right)  +1\right)
B\left(  r,E\right)  -B\left(  \bar{r},E\right)  \left(  1+D\left(  \bar
{r},R\right)  \right)  }{3B\left(  \bar{r},E\right)  \left(  1+D\left(
\bar{r},R\right)  \right)  -\left(  3D\left(  \bar{r},R\right)  +1\right)
B\left(  r,E\right)  } \label{p(r)Match}%
\end{equation}
and the radial pressure for the whole star is%
\begin{equation}
p_{r}(r)=\rho c^{2}\left\{
\begin{array}
[c]{cc}%
\frac{\left(  3D\left(  \bar{r},R\right)  +1\right)  B\left(  r,E\right)
-B\left(  \bar{r},E\right)  \left(  1+D\left(  \bar{r},R\right)  \right)
}{3B\left(  \bar{r},E\right)  \left(  1+D\left(  \bar{r},R\right)  \right)
-\left(  3D\left(  \bar{r},R\right)  +1\right)  B\left(  r,E\right)  } &
\bar{r}>r\geq0\\
D\left(  \bar{r},R\right)  & r=\bar{r}\\
\frac{\sqrt{3c^{2}-\kappa\rho r^{2}}-\sqrt{3c^{2}-\kappa\rho R^{2}}}%
{3\sqrt{3c^{2}-\kappa\rho R^{2}}-\sqrt{3c^{2}-\kappa\rho r^{2}}} & r>\bar{r}%
\end{array}
\right.  ,
\end{equation}
from which is possible to compute the pressure at the center of the star. One
finds%
\begin{equation}
p_{r}(0)=p_{c}=\rho c^{2}\frac{\left(  3D\left(  \bar{r},R\right)  +1\right)
\sqrt{3c^{2}g_{2}^{2}(E/E_{\mathrm{Pl}})}-B\left(  \bar{r},E\right)  \left(
1+D\left(  \bar{r},R\right)  \right)  }{3B\left(  \bar{r},E\right)  \left(
1+D\left(  \bar{r},R\right)  \right)  -\left(  3D\left(  \bar{r},R\right)
+1\right)  \sqrt{3c^{2}g_{2}^{2}(E/E_{\mathrm{Pl}})}} \label{pc}%
\end{equation}
and in order to have a finite $p_{c}$, we have to impose that the denominator
of $\left(  \ref{pc}\right)  $ be not nought, namely%
\begin{equation}
\frac{24c^{4}g_{2}^{2}(E/E_{\mathrm{Pl}})-9\kappa\rho c^{2}g_{2}%
^{2}(E/E_{\mathrm{Pl}})R^{2}}{9c^{2}\kappa\rho-3\kappa^{2}\rho^{2}R^{2}%
-\kappa\rho c^{2}g_{2}^{2}(E/E_{\mathrm{Pl}})}\neq\bar{r}^{2}. \label{Den}%
\end{equation}
Due to the complexity of the expression $\left(  \ref{p(r)Match}\right)  $, it
is useful to discuss the following limiting cases:

\begin{description}
\item[1)] $g_{2}(E/E_{\mathrm{Pl}})\rightarrow0$. Although the central
pressure $p_{c}$ approaches a finite and real limit%
\begin{equation}
p_{c}\simeq-\frac{\rho c^{2}}{3},
\end{equation}
the constant $A$ in $\left(  \ref{Match}\right)  $ becomes imaginary. Moreover
the inequality $\left(  \ref{BBm}\right)  $ becomes dominated by the $A\left(
E/E_{\mathrm{Pl}}\right)  $ function which is divergent allowing the
underlying mass to assume any value. For this reason, this limit will be discarded.

\item[2)] $g_{2}(E/E_{\mathrm{Pl}})\rightarrow\infty$. In this case, Eq.
$\left(  \ref{Den}\right)  $ becomes%
\begin{equation}
\frac{9\kappa\rho c^{2}R^{2}-24c^{4}}{\kappa\rho c^{2}}\neq\bar{r}^{2}%
\end{equation}
and by imposing%
\begin{equation}
R<\sqrt{\frac{c^{2}}{3\pi G\rho}},
\end{equation}
we obtain a Buchdahl-Bondi-like bound, because the mass density becomes%
\begin{equation}
\rho=M\left[  \frac{4\pi}{3}\left(  R^{3}-\bar{r}^{3}\right)  \right]  ^{-1}.
\label{rhoBar}%
\end{equation}
In this limit, the central pressure becomes%
\begin{equation}
p_{c}\simeq\rho c^{2}D\left(  \bar{r},R\right)  =\rho_{0}c^{2}\frac{\left(
\sqrt{3c^{2}-\kappa\rho_{0}\bar{r}^{2}}-\sqrt{3c^{2}-\kappa\rho_{0}R^{2}%
}\right)  }{\left(  3\sqrt{3c^{2}-\kappa\rho_{0}R^{2}}-\sqrt{3c^{2}-\kappa
\rho_{0}\bar{r}^{2}}\right)  }, \label{pcK}%
\end{equation}
from which is possible to obtain information on the radius of the star%
\begin{equation}
R=\sqrt{\frac{1}{\kappa\rho_{0}}}{{\frac{\sqrt{\left(  24{c}^{2}+\kappa
\rho_{0}\bar{r}^{2}\right)  p_{c}^{2}+\left(  12\rho_{0}{c}^{4}+2{c}^{2}%
\kappa\rho_{0}^{2}\bar{r}^{2}\right)  p_{c}+{c}^{4}\kappa\rho_{0}^{3}\,\bar
{r}^{2}}}{\,\rho_{0}{c}^{2}+3p_{c}}}.} \label{R}%
\end{equation}
Note that when $\bar{r}\rightarrow0$, we recover the usual Buchdahl-Bondi
bound. On the other hand, it is possible to have the expression of the
intermediate radius $\bar{r}$ as a function of $R,\rho_{0},\rho$ and $p_{c}$%
\begin{equation}
\bar{r}=\sqrt{\frac{1}{\kappa\rho_{0}}}{\frac{\sqrt{\left(  9\,\kappa\,{R}%
^{2}\rho_{0}-24\,{c}^{2}\right)  p_{c}^{2}+6\,{c}^{2}\rho_{0}\left(
\kappa\,{R}^{2}\rho_{0}-2\,{c}^{2}\right)  p_{c}+{R}^{2}{c}^{4}\rho_{0}%
^{3}\kappa}}{\rho_{0}{c}^{2}+p_{c}}}{.}%
\end{equation}

\end{description}

\subsubsection{Case b)}

In this case, the whole star is distorted by Gravity's Rainbow and the
boundary is set very close to the core. The integration of Eq.$\left(
\ref{TOVin}\right)  $ with the condition $p_{r}(R)=0$, leads to%
\begin{equation}
p_{r}(r)=\rho c^{2}\frac{\left(  \sqrt{3c^{2}g_{2}^{2}(E/E_{\mathrm{Pl}%
})-\kappa\rho r^{2}}-\sqrt{3c^{2}g_{2}^{2}(E/E_{\mathrm{Pl}})-\kappa\rho
R^{2}}\right)  }{\left(  3\sqrt{3c^{2}g_{2}^{2}(E/E_{\mathrm{Pl}})-\kappa\rho
R^{2}}-\sqrt{3c^{2}g_{2}^{2}(E/E_{\mathrm{Pl}})-\kappa\rho r^{2}}\right)
}.\label{p(r)}%
\end{equation}
Because of Eq.$\left(  \ref{eq:mg2}\right)  $ at the boundary $R$, we find%
\begin{equation}
\rho=g_{2}^{2}(E/E_{\mathrm{Pl}})\frac{3M}{4\pi R^{3}}=g_{2}^{2}%
(E/E_{\mathrm{Pl}})\tilde{\rho},\label{rel}%
\end{equation}
where $\tilde{\rho}$ is the mass density in ordinary GR. Thus Eq.$\left(
\ref{p(r)}\right)  $ becomes%
\begin{equation}
p_{r}(r)=g_{2}^{2}(E/E_{\mathrm{Pl}})\tilde{\rho}c^{2}\frac{\left(
\sqrt{3c^{2}-\kappa\tilde{\rho}r^{2}}-\sqrt{3c^{2}-\kappa\tilde{\rho}R^{2}%
}\right)  }{\left(  3\sqrt{3c^{2}-\kappa\tilde{\rho}R^{2}}-\sqrt{3c^{2}%
-\kappa\tilde{\rho}r^{2}}\right)  }=g_{2}^{2}(E/E_{\mathrm{Pl}})\tilde{p}%
_{r}(r).\label{p(r)1}%
\end{equation}
It is immediate to recognize that all the properties obtained in ordinary GR
are here valid, except for the pressure which scales with $g_{2}%
^{2}(E/E_{\mathrm{Pl}})$. The same behavior appears of course, when we
describe the pressure in terms of the mass $M$ and the radius $R$. Indeed,
always with the help of Eq.$\left(  \ref{eq:mg2}\right)  $, one gets%
\begin{equation}
p_{r}(r)=\frac{3Mg_{2}^{2}(E/E_{\mathrm{Pl}})}{4\pi R^{3}}c^{2}\frac
{\sqrt{c^{2}-2MGr^{2}/R^{3}}-\sqrt{c^{2}-2MG/R}}{3\sqrt{c^{2}-2MG/R}%
-\sqrt{c^{2}-2MGr^{2}/R^{3}}}=g_{2}^{2}(E/E_{\mathrm{Pl}})\tilde{p}%
_{r}(r)\label{p(r)M}%
\end{equation}
and the Buchdahl-Bondi bound is preserved. We can now compute the pressure at
the center of the star to obtain%
\begin{equation}
p_{r}(0)=p_{c}=g_{2}^{2}(E/E_{\mathrm{Pl}})\tilde{p}_{r}(0)=g_{2}%
^{2}(E/E_{\mathrm{Pl}})\tilde{\rho}c^{2}\frac{\left(  \sqrt{3c^{2}}%
-\sqrt{3c^{2}-\kappa\tilde{\rho}R^{2}}\right)  }{\left(  3\sqrt{3c^{2}%
-\kappa\tilde{\rho}R^{2}}-\sqrt{3c^{2}}\right)  }=g_{2}^{2}(E/E_{\mathrm{Pl}%
})\tilde{p}_{c}\label{p(0)}%
\end{equation}
while in terms of the mass $M$, we obtain%
\begin{equation}
p_{r}(0)=p_{c}=\frac{3Mg_{2}^{2}(E/E_{\mathrm{Pl}})}{4\pi R^{3}}c^{2}%
\frac{c-\sqrt{c^{2}-2MG/R}}{3\sqrt{c^{2}-2MG/R}-c}=g_{2}^{2}(E/E_{\mathrm{Pl}%
})\tilde{p}_{c}.\label{p(0)a}%
\end{equation}
Because of the pressure scaling, we find that the radius of the star can be
computed in the same way of the undeformed case. Indeed, in terms of the
rescaled density we find%
\begin{equation}
R=\,\sqrt{\frac{3c^{2}}{8\tilde{\rho}\pi G}\left[  1-\frac{\left(  \tilde
{\rho}{c}^{2}+\tilde{p}_{c}\right)  ^{2}}{\,\left(  \tilde{\rho}{c}%
^{2}+3\tilde{p}_{c}\right)  ^{2}}\right]  }.\label{Rrho}%
\end{equation}
The same undeformed result is obtained in terms of the mass $M$%
\begin{equation}
R=\frac{2MG\,}{c^{2}\sqrt{1-\frac{\left(  \tilde{\rho}\,{c}^{2}+\tilde{p}%
_{c}\right)  ^{2}}{\,\left(  \tilde{\rho}\,{c}^{2}+3\tilde{p}_{c}\right)
^{2}}}},\label{RM}%
\end{equation}
where we have used the Schwarzschild form on the boundary of the star.
However, when we go back to the deformed pressure and energy density, we find
that the undeformed radius $R$ described by $\left(  \ref{Rrho}\right)  $,
becomes\footnote{Note that the relation between the undeformed star radius $R$
and the deformed $\tilde{R}$ is%
\begin{equation}
R_{d}=R/g_{2}^{2/3}(E/E_{\mathrm{Pl}})
\end{equation}
as suggested by the expression $\left(  \ref{rel}\right)  $.}%
\begin{equation}
\frac{R}{g_{2}(E/E_{\mathrm{Pl}})}=\,\sqrt{\frac{3c^{2}}{8\rho\pi G}\left[
1-\frac{\left(  \rho{c}^{2}+p_{c}\right)  ^{2}}{\,\left(  \rho{c}^{2}%
+3p_{c}\right)  ^{2}}\right]  }.\label{RrhoG}%
\end{equation}
When $g_{2}(E/E_{\mathrm{Pl}})\gg1$, to obtain the shrinking of the radius of
the star $R$, necessarily we need $\tilde{\rho}\,{c}^{2}\gg3\tilde{p}_{c}$,
since the central pressure can be large but finite. When $R$ is small, we find%
\begin{equation}
p_{c}\simeq g_{2}^{2}(E/E_{\mathrm{Pl}})\frac{2\pi G\tilde{\rho}^{2}R^{2}}%
{3}+O\left(  R^{4}\right)
\end{equation}
or, in terms of the mass $M$,%
\begin{equation}
p_{c}\simeq g_{2}^{2}(E/E_{\mathrm{Pl}})\frac{3M^{2}G}{8\pi R^{4}}+O\left(
R^{4}\right)  .
\end{equation}
This also means that from Eq.$\left(  \ref{RM}\right)  $, $M$ must be small.
Notice that in terms of $\rho$ the equilibrium condition becomes
\begin{equation}
R<\frac{cg_{2}(E/E_{\mathrm{Pl}})}{\sqrt{3\pi G\rho}}.\label{boundrho}%
\end{equation}

In the standard framework $g_{2}(E/E_{\mathrm{Pl}})=1$ and Eq.(\ref{boundrho})
imply that when Planckian densities are approached, $\rho\approx\rho_{Pl}$,
one gets
\begin{equation}
R\lesssim l_{\mathrm{Pl}},
\end{equation}
i.e. only stars smaller than the Planck size can satisfy the TOV equilibrium equation. Instead,
in our Rainbow scenario, at Planckian densities we get
\begin{equation}
R\lesssim g_{2}(E/E_{\mathrm{Pl}})l_{\mathrm{Pl}},\label{RG_Pdensity}%
\end{equation}
suggesting that macroscopic stars are also allowed, if the function
$g_{2}(E/E_{\mathrm{Pl}})$ is very large.

\subsection{ The redshift function for the constant energy density case}

In the case of a constant density, the redshift function becomes%
\begin{equation}
\Phi(r)+K=-\int_{0}^{r}\frac{dp/dr^{\prime}}{\rho c^{2}+p(r^{\prime}%
)}dr^{\prime}.
\end{equation}
Because of the modification due to Gravity's Rainbow, we are forced to
separate the discussion of the redshift function into two cases. We begin with
the case a

\subsubsection{Case a}

In this case the computation of the redshift function separates into two
pieces%
\begin{equation}
\Phi(r)+K=-\int_{0}^{\bar{r}}\frac{dp/dr^{\prime}}{\rho c^{2}+p(r^{\prime}%
)}dr^{\prime}-\int_{\bar{r}}^{r}\frac{dp/dr^{\prime}}{\rho c^{2}+p(r^{\prime
})}dr^{\prime}=I_{1}+I_{2},\label{Phi(r)a}%
\end{equation}
where $\bar{r}$ has been defined in Eq.$\left(  \ref{dm}\right)  $ and the
related range in $\left(  \ref{radius}\right)  $. Plugging Eq.$\left(
\ref{TOVin}\right)  $ into the first integral one finds%
\begin{align}
I_{1} &  =-\int_{0}^{\bar{r}}\frac{dp/dr^{\prime}}{\rho c^{2}+p(r^{\prime}%
)}dr^{\prime}=-\int_{0}^{\bar{r}}\frac{dp/dr^{\prime}}{g_{2}^{2}%
(E/E_{\mathrm{Pl}})\tilde{\rho}c^{2}+p(r^{\prime})}dr^{\prime}\nonumber\\
&  =\frac{\kappa}{2c^{2}}\int_{0}^{\bar{r}}\frac{r^{\prime}\left(
3p(r^{\prime})+g_{2}^{2}(E/E_{\mathrm{Pl}})\tilde{\rho}c^{2}\right)  }{\left[
3c^{2}g_{2}^{2}(E/E_{\mathrm{Pl}})-\kappa g_{2}^{2}(E/E_{\mathrm{Pl}}%
)\tilde{\rho}r^{\prime2}\right]  }dr^{\prime}=I_{1a}+I_{1b},
\end{align}
where%
\begin{equation}
I_{1a}=\frac{3\kappa}{2c^{2}}\int_{0}^{\bar{r}}\frac{r^{\prime}p(r^{\prime}%
)}{\left[  3c^{2}g_{2}^{2}(E/E_{\mathrm{Pl}})-\kappa g_{2}^{2}%
(E/E_{\mathrm{Pl}})\tilde{\rho}r^{\prime2}\right]  }dr^{\prime}%
\end{equation}
and%
\begin{equation}
I_{1b}=\frac{\kappa\tilde{\rho}}{2}\int_{0}^{\bar{r}}\frac{r^{\prime
}dr^{\prime}}{\left[  3c^{2}-\kappa\tilde{\rho}r^{\prime2}\right]  }=-\frac
{1}{4}\ln\left(  \frac{3c^{2}-\kappa\tilde{\rho}\bar{r}^{2}}{3c^{2}}\right)  .
\end{equation}
Plugging Eq.$\left(  \ref{p(r)Match}\right)  $ into the integral $I_{1a}$, one
gets%
\begin{equation}
I_{1a}=\frac{3\kappa\tilde{\rho}}{2}\int_{0}^{\bar{r}}\frac{r^{\prime}\left(
3C\left(  \bar{r},R\right)  +1\right)  \tilde{B}\left(  r^{\prime},E\right)
-\tilde{B}\left(  \bar{r},E\right)  \left(  1+C\left(  \bar{r},R\right)
\right)  }{\left[  3c^{2}-\kappa\tilde{\rho}r^{\prime2}\right]  \left[
3\tilde{B}\left(  \bar{r},E\right)  \left(  1+C\left(  \bar{r},R\right)
\right)  -\left(  3C\left(  \bar{r},R\right)  +1\right)  \tilde{B}\left(
r^{\prime},E\right)  \right]  }dr^{\prime},
\end{equation}
where we have used the following relationship%
\begin{equation}
B\left(  r,E\right)  =\sqrt{3c^{2}g_{2}^{2}(E/E_{\mathrm{Pl}})-\kappa\rho
r^{2}}=g_{2}(E/E_{\mathrm{Pl}})\sqrt{3c^{2}-\kappa\tilde{\rho}r^{2}}%
=g_{2}(E/E_{\mathrm{Pl}})\tilde{B}\left(  r,E\right)  .
\end{equation}
Define the new variable%
\begin{equation}
3c^{2}-\kappa\tilde{\rho}r^{\prime2}=y^{2}\Longrightarrow-\kappa\tilde{\rho
}r^{\prime}dr^{\prime}=ydy,
\end{equation}
then $I_{1}$ becomes%
\begin{align}
I_{1a} &  =-\frac{3}{2}\int_{\sqrt{3c^{2}}}^{y\left(  \bar{r}\right)  }%
\frac{\left(  3C\left(  \bar{r},R\right)  +1\right)  y-\tilde{B}\left(
\bar{r},E\right)  \left(  1+C\left(  \bar{r},R\right)  \right)  }{y\left[
3\tilde{B}\left(  \bar{r},E\right)  \left(  1+C\left(  \bar{r},R\right)
\right)  -\left(  3C\left(  \bar{r},R\right)  +1\right)  y\right]
}dy\nonumber\\
&  =-\frac{3}{2}\int_{\sqrt{3c^{2}}}^{y\left(  \bar{r}\right)  }\frac
{C_{1}y-C_{2}}{y\left[  3C_{2}-C_{1}y\right]  }dy,
\end{align}
where%
\begin{align}
C_{1} &  =3C\left(  \bar{r},R\right)  +1\nonumber\\
C_{2} &  =\tilde{B}\left(  \bar{r},E\right)  \left(  1+C\left(  \bar
{r},R\right)  \right)  .
\end{align}
Now $I_{1a}$ can be easily integrated to give%
\begin{equation}
I_{1a}=-\frac{3}{2}\int_{\sqrt{3c^{2}}}^{y\left(  \bar{r}\right)  }\frac
{C_{1}y-C_{2}}{y\left[  3C_{2}-C_{1}y\right]  }dy=\ln\left(  \frac
{3C_{2}-C_{1}y\left(  \bar{r}\right)  }{3C_{2}-C_{1}\sqrt{3c^{2}}}\right)
+\frac{1}{2}\ln\left(  \frac{y\left(  \bar{r}\right)  }{\sqrt{3c^{2}}}\right)
\end{equation}
and%
\begin{equation}
I_{1}=\ln\left(  \frac{3C_{2}-C_{1}y\left(  \bar{r}\right)  }{3C_{2}%
-C_{1}\sqrt{3c^{2}}}\right)  .
\end{equation}
Following the same procedure for $I_{2}$, one gets%
\begin{align}
I_{2} &  =-\int_{\bar{r}}^{r}\frac{dp/dr^{\prime}}{\rho c^{2}+p(r^{\prime}%
)}dr^{\prime}=\frac{\kappa}{2c^{2}}\int_{\bar{r}}^{r}\frac{r^{\prime}\left(
3p(r^{\prime})+\rho c^{2}\right)  }{\left[  3c^{2}-\kappa\rho r^{\prime
2}\right]  }dr^{\prime}\nonumber\\
&  \underset{3c^{2}-\kappa\rho r^{\prime2}=z^{2}}{=}\ln\left(  \frac{3z\left(
R\right)  -z\left(  r\right)  }{3z\left(  R\right)  -z\left(  \bar{r}\right)
}\right)  .
\end{align}
Therefore Eq.$\left(  \ref{Phi(r)a}\right)  $ becomes%
\begin{equation}
\Phi(r)+K=\ln\left(  \frac{3C_{2}-C_{1}y\left(  \bar{r}\right)  }{3C_{2}%
-C_{1}\sqrt{3c^{2}}}\right)  +\ln\left(  \frac{3z\left(  R\right)  -z\left(
r\right)  }{3z\left(  R\right)  -z\left(  \bar{r}\right)  }\right)
=\ln\left(  \frac{\left(  3C_{2}-C_{1}y\left(  \bar{r}\right)  \right)
\left(  3z\left(  R\right)  -z\left(  r\right)  \right)  }{\left(
3C_{2}-C_{1}\sqrt{3c^{2}}\right)  \left(  3z\left(  R\right)  -z\left(
\bar{r}\right)  \right)  }\right)  .
\end{equation}
At the boundary of the star we obtain
\begin{gather}
\exp2\left(  \Phi(R)+K\right)  =\left(  \frac{\left(  3C_{2}-C_{1}y\left(
\bar{r}\right)  \right)  \left(  2z\left(  R\right)  \right)  }{\left(
3C_{2}-C_{1}\sqrt{3c^{2}}\right)  \left(  3z\left(  R\right)  -z\left(
\bar{r}\right)  \right)  }\right)  ^{2}\nonumber\\
\Longrightarrow\exp2K=\frac{1}{\exp2\Phi(R)}\left(  \frac{\left(  3C_{2}%
-C_{1}y\left(  \bar{r}\right)  \right)  \left(  2z\left(  R\right)  \right)
}{\left(  3C_{2}-C_{1}\sqrt{3c^{2}}\right)  \left(  3z\left(  R\right)
-z\left(  \bar{r}\right)  \right)  }\right)  ^{2},
\end{gather}
then%
\begin{equation}
\exp2\left(  \Phi(R)+K\right)  =\left(  \frac{\left(  3C_{2}-C_{1}y\left(
\bar{r}\right)  \right)  \left(  2z\left(  R\right)  \right)  }{\left(
3C_{2}-C_{1}\sqrt{3c^{2}}\right)  \left(  3z\left(  R\right)  -z\left(
\bar{r}\right)  \right)  }\right)  ^{2}%
\end{equation}
and
\begin{equation}
\exp2\Phi(r)=\exp2\Phi(R)\left(  \frac{\left(  3z\left(  R\right)  -z\left(
r\right)  \right)  }{\left(  2\sqrt{3c^{2}-\kappa\rho R^{2}}\right)  }\right)
^{2}.
\end{equation}
However, because of the Schwarzschild boundary condition, namely%
\begin{equation}
\exp2\Phi(R)=1-\frac{2MG}{c^{2}R},\label{Phi(R)}%
\end{equation}
and because of the Eq.$\left(  \ref{rhoM}\right)  $, one finds that the
redshift surface becomes%
\begin{equation}
\exp2\Phi(r)=\left(  1-\frac{2MG}{c^{2}R}\right)  \left(  \frac{\left(
3\sqrt{1-\frac{2MGR^{2}}{c^{2}\tilde{R}^{3}}}-\sqrt{1-\frac{2MGr^{2}}%
{c^{2}\tilde{R}^{3}}}\right)  }{\left(  2\sqrt{1-\frac{2MGR^{2}}{c^{2}%
\tilde{R}^{3}}}\right)  }\right)  ^{2},
\end{equation}
where we have used Eq.$\left(  \ref{Rtilde}\right)  $. In any case, on the
star surface the redshift factor reduces to
\begin{equation}
z=\frac{\triangle\lambda}{\lambda}=\frac{g_{1}(E/E_{\mathrm{Pl}})}{\exp
[\Phi(R)]}-1=\frac{g_{1}(E/E_{\mathrm{Pl}})}{\sqrt{1-\frac{2MG}{Rc^{2}}}}-1.
\end{equation}
The rainbow upper bound on the redshift factor
\begin{equation}
z\leq z_{\max}=3g_{1}(E/E_{\mathrm{Pl}})-1
\end{equation}
becomes $z_{\max}=2$ in the undeformed limit $g_{1}(E/E_{P})=1$, as expected.
It is clear that for energies comparable with $E_{\mathrm{Pl}}$, one can have
deviations from the usual redshift factor. Indeed, from%
\begin{equation}
g_{1}(E/E_{P})\simeq1+\alpha\frac{E}{E_{p}}+O\left(  \left(  \frac{E}{E_{p}%
}\right)  ^{2}\right)  ,
\end{equation}
where%
\begin{equation}
\alpha=\left(  \frac{dg_{1}(E/E_{P})}{dE}\right)  _{|E=0}\frac{1}{E_{p}%
},\label{alpha}%
\end{equation}
we have%
\begin{equation}
z_{\max}=2+3\alpha\frac{E}{E_{p}}+O\left(  \frac{E}{E_{p}}\right)  ^{2},
\end{equation}
with $\alpha$ $\lessgtr0$.

\subsubsection{Case b}

In the case of a constant density one can also calculate the redshift function
explicitly. Indeed, from Eq.$\left(  \ref{EqEq}\right)  $, we find%
\begin{equation}
\Phi(r)+K=-\int_{0}^{r}\frac{dp/dr^{\prime}}{g_{2}^{2}(E/E_{\mathrm{Pl}%
})\tilde{\rho}c^{2}+p(r^{\prime})}dr^{\prime} \label{Phi(r)}%
\end{equation}
and with the help of Eq.$\left(  \ref{p(r)M}\right)  $, one can write
\begin{equation}
p_{r}(y^{\prime})\underset{y^{\prime2}=c^{2}-2MGr^{\prime2}/R^{3}}{=}g_{2}%
^{2}(E/E_{\mathrm{Pl}})\tilde{\rho}c^{2}\frac{\left(  y^{\prime}-\sqrt
{c^{2}-2MG/R}\right)  }{\left(  3\sqrt{c^{2}-2MG/R}-y^{\prime}\right)  }.
\end{equation}
Thus%
\begin{equation}
\frac{dp_{r}}{dy^{\prime}}=g_{2}^{2}(E/E_{\mathrm{Pl}})\tilde{\rho}c^{2}%
\frac{2\sqrt{c^{2}-2MG/R}}{\left(  3\sqrt{c^{2}-2MG/R}-y^{\prime}\right)
^{2}}%
\end{equation}
and Eq.$\left(  \ref{Phi(r)}\right)  $ becomes
\begin{gather}
\Phi(y)+K=-\int_{c^{2}}^{y}\frac{dp/dy^{\prime}}{g_{2}^{2}(E/E_{\mathrm{Pl}%
})\tilde{\rho}c^{2}+p(y^{\prime})}dy^{\prime}\nonumber\\
=-\int_{c^{2}}^{y}\frac{dy^{\prime}}{\left(  3\sqrt{c^{2}-2MG/R}-y^{\prime
}\right)  }=\ln\left(  \frac{3\sqrt{c^{2}-2MG/R}-y}{3\sqrt{c^{2}-2MG/R}-c^{2}%
}\right)  .
\end{gather}
At the boundary of the star, we obtain $\exp2\Phi(R)=1-2MG/c^{2}R$, thus
\begin{gather}
\exp2\left(  \Phi(R)+K\right)  =\left(  \frac{3\sqrt{c^{2}-2MG/R}-y\left(
R\right)  }{3\sqrt{c^{2}-2MG/R}-c^{2}}\right)  ^{2}\nonumber\\
\Longrightarrow\exp2K=\frac{1}{\exp2\Phi(R)}\left(  \frac{3\sqrt{c^{2}%
-2MG/R}-y\left(  R\right)  }{3\sqrt{c^{2}-2MG/R}-c^{2}}\right)  ^{2},
\end{gather}
then
\begin{align}
\exp2\Phi(r)  &  =\exp2\Phi(R)\left(  \frac{3\sqrt{c^{2}-2MG/R}-\sqrt
{c^{2}-2MGr^{2}/R^{3}}}{2\sqrt{c^{2}-2MG/R}}\right)  ^{2}\\
&  =\exp\left(  \frac{1}{4c^{2}}\left(  3\sqrt{c^{2}-2MG/R}-\sqrt
{c^{2}-2MGr^{2}/R^{3}}\right)  ^{2}\right)  .
\end{align}
Explicitly%

\begin{equation}
\Phi(r)=\ln\left[  \frac{1}{4}\left(  3\sqrt{1-2MG/c^{2}R}-\sqrt
{1-2MGr^{2}/c^{2}R^{3}}\right)  \right]  \qquad r\in\left[  0,R\right]  .
\end{equation}
It is immediate to recognize that the behavior of the surface redshift is the
same of the case a), except for the range which here is related to the whole star.

\section{The Isotropic TOV equation and the EoS: variable energy density case}

\label{p4}In this section, we will consider an energy density profile of the
following form
\begin{equation}
\rho=Ar^{\alpha},\label{rhoV}%
\end{equation}
where $A$ is a constant with dimensions of an energy density divided by a
(length)$^{\alpha}$ with $\alpha\in\mathbb{R}$ to be determined. Solving
Eq.$\left(  \ref{m'(r)}\right)  $ leads to
\begin{equation}
m(r)=\int_{0}^{r}\frac{4\pi A}{g_{2}^{2}(E/E_{\mathrm{Pl}})}r^{\prime2+\alpha
}dr^{\prime}=\frac{4\pi A}{g_{2}^{2}(E/E_{\mathrm{Pl}})\left(  3+\alpha
\right)  }r^{3+\alpha}.\label{b(r)}%
\end{equation}
Plugging $\left(  \ref{rhoV}\right)  $ and $\left(  \ref{b(r)}\right)  $ into
Eq.$\left(  \ref{TOV}\right)  $, one finds
\begin{gather}
\omega\frac{d\rho\left(  r\right)  }{dr}=-\rho\left(  r\right)  \left(
\frac{c^{2}+\omega}{c^{2}}\right)  \frac{4\pi Gr^{3}\omega\rho(r)+Gm(r)c^{2}%
g_{2}^{2}(E/E_{\mathrm{Pl}})}{r^{2}\left[  1-2Gm(r)/rc^{2}\right]  c^{2}%
g_{2}^{2}(E/E_{\mathrm{Pl}})}\nonumber\\
\Downarrow\\
\alpha=-\left(  \frac{c^{2}+\omega}{\omega c^{2}}\right)  \frac{4\pi
GAr^{2+\alpha}\left(  \left(  3+\alpha\right)  \omega+c^{2}\right)  }{\left[
c^{2}g_{2}^{2}(E/E_{\mathrm{Pl}})\left(  3+\alpha\right)  -8\pi GAr^{2+\alpha
}\right]  }.
\end{gather}
It is immediate to see that $\forall\alpha\neq-2$, there is a singularity into
the TOV equation and a dependence on $r$ still persists. Therefore if we fix
$\alpha=-2$, one gets the relationship
\begin{equation}
1=\frac{3\left(  c^{2}+\omega\right)  ^{2}}{4\omega\left[  7c^{2}g_{2}%
^{2}(E/E_{\mathrm{Pl}})-3\right]  },\label{omega}%
\end{equation}
where we have set $A=3c^{2}/\left(  56\pi G\right)  $. We find an identity
when $\omega=1/3$, $\omega=3$, $c=1$ and $g_{2}(E/E_{\mathrm{Pl}})=1$.
Therefore in ordinary GR, TOV is satisfied for
\begin{equation}
p_{r}=\omega\rho\left(  r\right)  =\omega\frac{3c^{2}}{56\pi Gr^{2}}%
\end{equation}
and
\begin{equation}
m(r)=\frac{3c^{2}r}{14G}.
\end{equation}
The energy density in $\left(  \ref{MZ}\right)  $ has been found for the first
time by Misner and Zapolsky \cite{MisnerZapolsky}. When Gravity's Rainbow
comes into play, one can find the values of $\omega$ satisfying the constraint
$\left(  \ref{omega}\right)  $. One finds
\begin{equation}
\omega_{\pm}=\frac{14}{3}c^{2}g_{2}^{2}(E/E_{\mathrm{Pl}})-c^{2}-2\pm\frac
{2}{3}\sqrt{\Delta},\label{omsols}%
\end{equation}
where
\begin{equation}
\Delta=49c^{4}g_{2}^{4}(E/E_{\mathrm{Pl}})-21c^{4}g_{2}^{2}(E/E_{\mathrm{Pl}%
})-42c^{2}g_{2}^{2}(E/E_{\mathrm{Pl}})+9c^{2}+9.
\end{equation}
When $g_{2}(E/E_{\mathrm{Pl}})\gg1$, the asymptotic form of $\omega_{\pm}$ is
\begin{equation}
\omega_{+}\simeq\frac{28}{3}c^{2}g_{2}^{2}(E/E_{\mathrm{Pl}})-2c^{2}%
-4-\frac{3c^{2}}{28g_{2}^{2}(E/E_{\mathrm{Pl}})}+O\left(  \frac{1}{g^{4}%
}\right)  \simeq\frac{28}{3}c^{2}g_{2}^{2}(E/E_{\mathrm{Pl}})
\end{equation}
and
\begin{equation}
\omega_{-}\simeq\frac{3c^{2}}{28g_{2}^{2}(E/E_{\mathrm{Pl}})}+O\left(
\frac{1}{g^{4}}\right)  .
\end{equation}
It is immediate to see that both solutions acquire a dependence on
$g_{2}(E/E_{\mathrm{Pl}})$ which is decreasing for $\omega_{-}$ and increasing
for $\omega_{+}$. Note that at this stage, $E$ acts as a parameter independent
on the radial coordinate $r$. Of course, it is always possible to consider the
situation in which $g_{1}(E/E_{\mathrm{Pl}})\equiv g_{1}(E\left(  r\right)
/E_{\mathrm{Pl}})$ and $g_{2}(E/E_{\mathrm{Pl}})\equiv g_{2}(E\left(
r\right)  /E_{\mathrm{Pl}})$\cite{RGES}. However, this goes beyond the purpose
of this paper and it will be investigated elsewhere. Note that as in the
original model of Dev and Gleiser, $p_{r}\left(  R\right)  =0$, only if we
allow anisotropy. However, if we take under consideration the relation with
$\omega_{-}$, one can consider the situation in which
\begin{equation}
p_{r}\left(  R\right)  =\omega_{-}\rho\left(  R\right)  =\frac{9c^{4}}{1568\pi
Gg_{2}^{2}(E/E_{\mathrm{Pl}})R^{2}}\rightarrow0\label{p(r)MZ}%
\end{equation}
when $g_{2}(E/E_{\mathrm{Pl}})\gg1$ without invoking a boundary that goes to
infinity. As we can see, in this regime, the star seems to behave as dust,
because $\omega_{-}\rightarrow0$. For completeness, we present also the
expansion for small energies where $g_{1}(E/E_{\mathrm{Pl}})\simeq
g_{2}(E/E_{\mathrm{Pl}})\simeq1$. For example we can write for $\omega_{+}$
\begin{align}
\omega_{+} &  \simeq-{c}^{2}+{\frac{8}{3}}+{\frac{4}{3}\sqrt{4-3{c}^{2}}%
}+{\frac{7{\left(  \left(  3{c}^{2}-8\right)  \sqrt{4-3{c}^{2}}+12{c}%
^{2}-16\right)  }}{9{c}^{2}-12}}\left(  g_{2}(E/E_{\mathrm{Pl}})-1\right)
+O\left(  \left(  g-1\right)  ^{2}\right)  \nonumber\\
&  =-{c}^{2}+{\frac{8}{3}}+{\frac{4}{3}\sqrt{4-3{c}^{2}}}+{\frac{7{\left(
\left(  3{c}^{2}-8\right)  \sqrt{4-3{c}^{2}}+12{c}^{2}-16\right)  }}{9{c}%
^{2}-12}}\beta+O\left(  \beta^{2}\right)  \nonumber\\
&  \underset{{c}^{2}\rightarrow1}{=}3+21\beta+O\left(  \beta^{2}\right)
\end{align}
and for $\omega_{-}$
\begin{align}
\omega_{-} &  \simeq-{c}^{2}+{\frac{8}{3}}-{\frac{4}{3}\sqrt{4-3{c}^{2}}%
}-{\frac{7{\left(  \left(  3{c}^{2}-8\right)  \sqrt{4-3{c}^{2}}-12{c}%
^{2}+16\right)  }}{9{c}^{2}-12}}\left(  g_{2}(E/E_{\mathrm{Pl}})-1\right)
+O\left(  \left(  g-1\right)  ^{2}\right)  \nonumber\\
&  =-{c}^{2}+{\frac{8}{3}}-{\frac{4}{3}\sqrt{4-3{c}^{2}}}-{\frac{7{\left(
\left(  3{c}^{2}-8\right)  \sqrt{4-3{c}^{2}}-12{c}^{2}+16\right)  }}{9{c}%
^{2}-12}}\beta+O\left(  \beta^{2}\right)  \nonumber\\
&  \underset{{c}^{2}\rightarrow1}{=}{\frac{1}{3}}-{\frac{7}{3}}\beta+O\left(
\beta^{2}\right)  ,
\end{align}
where we have defined%
\begin{equation}
\beta=\left(  \frac{dg_{2}(E/E_{P})}{dE}\right)  _{|E=0}\frac{1}{E_{p}%
},\label{beta}%
\end{equation}
in analogy with definition $\left(  \ref{alpha}\right)  $. As regards the star
mass, one can easily verify that
\begin{equation}
m(r)=\frac{3c^{2}r}{g_{2}^{2}(E/E_{\mathrm{Pl}})14G}%
\end{equation}
and at the boundary $R$, one gets
\begin{equation}
M=m(R)=\frac{3c^{2}R}{g_{2}^{2}(E/E_{\mathrm{Pl}})14G}.\label{m(R)}%
\end{equation}

\subsection{The redshift function for the variable energy density case}

\label{p4a}The mass of the star at the boundary $R$, Eq.$\left(
\ref{m(R)}\right)  $, is useful also to determine the redshift factor. Indeed,
if we define the compactness of the star as%
\begin{equation}
\frac{MG}{Rc^{2}}=\frac{3}{g_{2}^{2}(E/E_{\mathrm{Pl}})14},
\end{equation}
then the surface redshift $z$ corresponding to the above compactness factor is
obtained as%
\begin{equation}
z=\frac{g_{1}(E/E_{\mathrm{Pl}})}{\sqrt{1-2MG/Rc^{2}}}-1=g_{1}%
(E/E_{\mathrm{Pl}})\left(  1-\frac{3}{7g_{2}^{2}(E/E_{\mathrm{Pl}})}\right)
^{-\frac{1}{2}}-1.
\end{equation}
It is immediate to see that only the case in which $g_{2}(E/E_{\mathrm{Pl}%
})>\sqrt{3/7}$ is allowed, otherwise $z$ would become imaginary. This means
that for an energy density profile of the form $\left(  \ref{rhoV}\right)  $,
the case in which $g_{2}(E/E_{\mathrm{Pl}})\leq\sqrt{3/7}$ is automatically
excluded. Moreover, if $g_{2}(E/E_{\mathrm{Pl}})$ is very large, we get%
\begin{equation}
z\simeq\frac{3g_{1}(E/E_{\mathrm{Pl}})}{14g_{2}^{2}(E/E_{\mathrm{Pl}})}.
\end{equation}
Note that when $g_{1}(E/E_{\mathrm{Pl}})\propto g_{2}^{2}(E/E_{\mathrm{Pl}})$,
then $z$ is approximately a constant. On the other hand, when we consider the
situation in which $E\ll E_{\mathrm{Pl}}$, one can have small deviations from
the undeformed redshift factor $z^{\ast}=\sqrt{7}/2-1\simeq0.322\,88$. Indeed
one finds%
\begin{equation}
z\simeq g_{1}(E/E_{\mathrm{Pl}})\frac{\sqrt{7}}{2}\left(  1-\frac{3}{8}%
\beta\frac{E}{E_{\mathrm{Pl}}}\right)  -1\simeq\left(  1+\alpha\frac{E}{E_{p}%
}\right)  \frac{\sqrt{7}}{2}\left(  1-\frac{3}{8}\beta\frac{E}{E_{\mathrm{Pl}%
}}\right)  -1\simeq z^{\ast}+\frac{\sqrt{7}E}{2E_{p}}\left(  \alpha-\frac
{3}{8}\beta\right)  ,
\end{equation}
with $\left(  \alpha-\frac{3}{8}\beta\right)  $ $\lessgtr0$, where we have
used definitions $\left(  \ref{alpha}\right)  $ and $\left(  \ref{beta}%
\right)  $.

\subsection{The redshift function for the Dev-Gleiser energy density case}

The combination of the constant and variable energy density profile considered
in section \ref{p3} and \ref{p4}, is known as the Dev-Gleiser\cite{DevGleiser}
energy density profile whose expression is%
\begin{equation}
\rho\left(  r\right)  =\rho+\frac{A}{r^{2}}, \label{DG}%
\end{equation}
where we have set $A=3c^{2}/\left(  56\pi G\right)  $. We know that in
ordinary GR, Dev-Gleiser solved the TOV equation in presence of anisotropy
showing that the pressureless condition on the boundary could be satisfied.
However in the isotropic case, it is not trivial to find solutions for the TOV
equation. Nevertheless, it is again possible to discuss the behavior of the
redshift for such a configuration. Indeed, it is immediate to see that
Eq.$\left(  \ref{m'(r)}\right)  $ can be easily solved to give%
\begin{equation}
m\left(  r\right)  =\int_{0}^{r}\frac{4\pi\rho(r^{\prime})r^{\prime2}}%
{g_{2}^{2}(E/E_{\mathrm{Pl}})}dr^{\prime}=\frac{4\pi}{g_{2}^{2}%
(E/E_{\mathrm{Pl}})}\left(  \frac{\rho r^{3}}{3}+Ar\right)
\end{equation}
and the total mass $M$ for a star of radius $R$ is simply%
\begin{equation}
M=\frac{4\pi}{g_{2}^{2}(E/E_{\mathrm{Pl}})}\left(  \frac{\rho R^{3}}%
{3}+AR\right)  .
\end{equation}
To simplify the computation we have considered the case b) of section \ref{p3}
where $R\simeq\alpha l_{\mathrm{Pl}}$. Then, we can define the compactness of
the star as%
\begin{equation}
\frac{MG}{Rc^{2}}=\frac{4\pi}{g_{2}^{2}(E/E_{\mathrm{Pl}})c^{2}}\left(
\frac{\rho R^{2}}{3}+A\right)  ,
\end{equation}
the surface redshift $z$ corresponding to the above compactness factor is
obtained as%
\begin{equation}
z=\frac{g_{1}(E/E_{\mathrm{Pl}})}{\sqrt{1-2MG/Rc^{2}}}-1=g_{1}%
(E/E_{\mathrm{Pl}})\left(  1-\frac{8\pi}{g_{2}^{2}(E/E_{\mathrm{Pl}})c^{2}%
}\left(  \frac{\rho R^{2}}{3}+A\right)  \right)  ^{-\frac{1}{2}}-1.
\end{equation}
Even in the Dev-Gleiser profile only the case in which $g_{2}(E/E_{\mathrm{Pl}%
})\gg1$ is allowed, otherwise $z$ would become imaginary. This means that for
an energy density profile of the form $\left(  \ref{DG}\right)  $, the case in
which $g_{2}(E/E_{\mathrm{Pl}})\ll1$ is automatically excluded. Instead, if
$g_{2}(E/E_{\mathrm{Pl}})$ is very large, we get%
\begin{equation}
z\simeq\frac{4\pi g_{1}(E/E_{\mathrm{Pl}})}{g_{2}^{2}(E/E_{\mathrm{Pl}})c^{2}%
}\left(  \frac{\rho R^{2}}{3}+A\right)  .
\end{equation}
It is immediate to see that even if $g_{1}(E/E_{\mathrm{Pl}})\propto g_{2}%
^{2}(E/E_{\mathrm{Pl}})$, then $z$ cannot be approximated by a constant as in
the previous subsection, because a dependence on the radius of the star $R$
still persists, not having found, for the Dev-Gleiser energy-density profile, a simple
analytical expression analogous to (\ref{ineq}).

\section{Conclusions}

\label{p5}In this paper we have considered the effects of Gravity's Rainbow on
the TOV equations. After having derived the deformed TOV equations, we have
focused our attention on two particular simple cases: the constant energy
density profile and the variable energy density profile, respectively. Since
the deformation induced by Gravity's Rainbow is expected to become more
relevant when Planckian energy density is approached, we have considered two
specific situations for the constant energy density profile: the first one
deals with a star which has a deformed core and an undeformed external region,
that's to say a two-fluid model. The second one considers a star which is
deformed everywhere. Even if it is possible to compute a pressure for the
whole star in both situations, due to the complexity of the analytical
expressions, we have considered two limiting cases: $g_{2}(E/E_{\mathrm{Pl}%
})\rightarrow\infty$ and $g_{2}(E/E_{\mathrm{Pl}})\rightarrow0$. For the
two-fluid model or case \textbf{a)} of section \ref{p3}, only the
$g_{2}(E/E_{\mathrm{Pl}})\rightarrow\infty$ limit has been considered to avoid
complex pressures and infinite masses. In this extreme limit, one finds that
the central pressure depends on the undeformed mass density and on the
boundary $\bar{r}$ where Gravity's Rainbow switches off, namely the core is
cut off as shown in Eq.$\left(  \ref{rhoBar}\right)  $. It is clear that this
is the result of a crude approximation and the addition of a dependence on the
radius $r$ from $g_{1}(E/E_{\mathrm{Pl}})\equiv g_{1}(E\left(  r\right)
/E_{\mathrm{Pl}})$ and $g_{2}(E/E_{\mathrm{Pl}})\equiv g_{2}(E\left(
r\right)  /E_{\mathrm{Pl}})$\cite{RGES} could give light to this result. On
the other hand, when Gravity's Rainbow is applied to the whole star or case
\textbf{b)}, we find that the star can survive in the TOV sense and that, due
to the $g_{2}$ factor, the size on the star does not necessarily become
Planckian (Eq.$\left(  \ref{RG_Pdensity}\right)  $). Even in this case, we do
not know if some corrections due to a full quantum gravitational theory can
corroborate or destroy the picture. Regarding the redshift factor for both
cases \textbf{a)} and \textbf{b)}, we find that the deformation is induced by
$g_{1}(E/E_{\mathrm{Pl}})$ only and there is a deviation that could be
detected in principle, even for small values of $E$. As regards, the variable
energy density profile, we have found that the parameter of the EoS $\omega$
cannot be considered as constant but acquires a dependence on
$E/E_{\mathrm{Pl}}$. Even for the variable case, we have considered the
$g_{2}(E/E_{\mathrm{Pl}})\rightarrow\infty$ limit, to avoid infinite masses.
In this regime, we have found two solutions $\omega_{\pm}$: $\omega_{+}$ is
divergent when $g_{2}(E/E_{\mathrm{Pl}})\rightarrow\infty$, while $\omega
_{-}\rightarrow0$, when in the same limit. While $\omega_{+}$ must be
discarded , we can see that $\omega_{-}$ can represent a form of
\textquotedblleft\textit{Gravity's Rainbow dust}\textquotedblright. It is
interesting to note that the vanishing of the pressure at the boundary $R$ is
here reached as a limit procedure. Indeed as shown by Dev and
Gleiser\cite{DevGleiser}, only if we introduce anisotropy, we can have the
exact vanishing of the pressure at the boundary. Regarding the redshift we
here find that $z$ depends on both the Rainbow's functions. As a particular
case, one can fix the ideas where $g_{1}(E/E_{\mathrm{Pl}})\propto g_{2}%
^{2}(E/E_{\mathrm{Pl}})$. With this choice, one finds that the redshift factor
is almost constant. Almost because, the exact value $z=3/14$ is reached when
$g_{1}(E/E_{\mathrm{Pl}})=g_{2}^{2}(E/E_{\mathrm{Pl}})$ and not simply
proportional. The same situation appears also for the Dev-Gleiser potential,
where we have only considered the redshift problem since the pressure
computation needs a more elaborate scheme. In summary, it seems that the
distortion created by Gravity's Rainbow on the TOV equation is able to create
stars that are really Planckian in density without necessarily being Planckian
in size. These \textquotedblleft\textit{Planck stars}\textquotedblright\ seem
to be completely different by the Planck stars proposed by Rovelli and
Vidotto\cite{Rovelli:2014cta}. Indeed, for an appropriate choice of the
function $g_{2}(E/E_{\mathrm{Pl}})$, the Buchdahl-Bondi bound is satisfied and
the collapse never appears. It is clear that the correction due to a
dependence on the radial coordinate of the form $g_{1}(E/E_{\mathrm{Pl}%
})\equiv g_{1}(E\left(  r\right)  /E_{\mathrm{Pl}})$ and $g_{2}%
(E/E_{\mathrm{Pl}})\equiv g_{2}(E\left(  r\right)  /E_{\mathrm{Pl}}%
)$ or a correction induced by a quantum gravitational calculation
could considerably improve the present stage of the computation.

\appendix{}

\section{Derivation of the TOV\ Equations in Gravity's Rainbow}

\label{App1}For a static fluid, we can define
\begin{equation}
u^{1}=\frac{dr}{d\tau}=0\quad u^{2}=\frac{d\theta}{d\tau}=0\quad u^{3}%
=\frac{d\phi}{d\tau}=0
\end{equation}
and with the help of the normalization $u_{\mu}u^{\mu}=-1$, we can write
\begin{equation}
-1=-\frac{e^{2\Phi(r)}}{g_{1}^{2}(E/E_{\mathrm{Pl}})}u^{0}u^{0}\rightarrow
\quad u^{0}=\frac{dt}{d\tau}=g_{1}(E/E_{\mathrm{Pl}})e^{-\Phi(r)}.
\end{equation}
For the energy-momentum stress tensor, one finds%
\begin{align}
T^{00}  &  =\rho(r)c^{2}g_{1}^{2}(E/E_{\mathrm{Pl}})c^{-2}e^{-2\Phi
(r)}\nonumber\\
T^{11}  &  =g_{2}^{2}(E/E_{\mathrm{Pl}})p\left(  1-b(r)/r\right) \nonumber\\
T^{22}  &  =g_{2}^{2}(E/E_{\mathrm{Pl}})pr^{-2}\nonumber\\
T^{33}  &  =g_{2}^{2}(E/E_{\mathrm{Pl}})pr^{-2}\sin^{-2}\theta,
\end{align}
and in terms of the mixed tensor, one gets
\begin{equation}
T_{0}^{0}=-\rho(r)c^{2}\qquad T_{1}^{1}=T_{2}^{2}=T_{2}^{2}=p(r).
\end{equation}
Thus from Einstein's equations ($\kappa=8\pi G$) we obtain
\begin{equation}
G_{00}=\kappa T_{00}\qquad\rightarrow\qquad b^{\prime}(r)=\frac{\kappa
\rho(r)c^{2}r^{2}}{c^{4}g_{2}^{2}(E/E_{\mathrm{Pl}})}%
\end{equation}
and
\begin{equation}
G_{11}=\kappa T_{11}\qquad\rightarrow\qquad\Phi^{\prime}(r)=\frac{\kappa
r^{3}p(r)/c^{2}g_{2}^{2}(E/E_{\mathrm{Pl}})+2Gm(r)}{2r^{2}c^{2}\left[
1-\frac{2Gm(r)}{rc^{2}}\right]  }. \label{eq:Pot-1}%
\end{equation}
From the conservation of the stress-energy tensor $T_{;\nu}^{\mu\nu}=0$
follows%
\[
T_{;\nu}^{\mu\nu}=\frac{\partial T^{\mu\nu}}{\partial x^{\nu}}+\Gamma
_{\beta\nu}^{\mu}T^{\beta\nu}+\Gamma_{\nu\beta}^{\nu}T^{\mu\beta}=0.
\]
However, for practical purposes, it is convenient to adopt the mixed
stress-energy tensor leading to
\begin{align}
\mu &  =0\qquad\Longrightarrow\qquad{\frac{\partial T_{0}^{0}\left(
t,r,\theta,\phi\right)  }{\partial t}}=0,\nonumber\\
\mu &  =2\qquad\Longrightarrow\qquad{\frac{\partial T_{0}^{0}\left(
t,r,\theta,\phi\right)  }{\partial\theta}}=0\nonumber\\
\mu &  =3\qquad\Longrightarrow\qquad{\frac{\partial T_{0}^{0}\left(
t,r,\theta,\phi\right)  }{\partial\phi}}=0
\end{align}
and%
\begin{equation}
\mu=1\qquad\Longrightarrow\qquad{\frac{\partial p(r)}{\partial r}+}%
\Phi^{\prime}\left(  r\right)  \left(  \rho(r)c^{2}+p(r)\right)  =0.
\label{Tmn}%
\end{equation}

\section{The Dev-Gleiser Energy Density Profile Induced by the ZPE in a
Gravity's Rainbow Context}

In this Section we shall consider the formalism outlined in detail in Ref.
\cite{RGGM,RGGM1}, where the graviton one loop contribution to a fixed
background is used. The latter contribution is evaluated through a variational
approach with Gaussian trial wave functionals, and the divergences are taken
under control with the help of Gravity's Rainbow. We refer the reader to
Ref.\cite{RGGM,RGGM1} for details. In ordinary gravity the computation of ZPE
for quantum fluctuations of the \textit{pure gravitational field} can be
extracted by rewriting the Wheeler-DeWitt equation (WDW)\cite{DeWitt} in a
form which looks like an expectation value computation\cite{Garattini:2004zu, Garattini:2005kx, Garattini:2005ky}. We remind
the reader that the WDW equation is the quantum version of the classical
constraint which guarantees the invariance under time reparametrization. Its
original form with the cosmological term included is described by
\begin{equation}
N\left(  r\right)  \rightarrow e^{2\Phi(r)}\qquad\mathrm{and}\qquad b\left(
r\right)  \rightarrow\frac{2Gm(r)}{c^{2}}.
\end{equation}%
\begin{equation}
\mathcal{H}\Psi=\left[  \left(  2\kappa\right)  G_{ijkl}\pi^{ij}\pi^{kl}%
-\frac{\sqrt{g}}{2\kappa}\!{}\!\left(  \,\!^{3}R-2\Lambda\right)  \right]
\Psi=0. \label{WDW}%
\end{equation}
Note that $\mathcal{H}=0$ represents one of the classical constraints. The
other one is the invariance by spatial diffeomorphism. If we multiply
Eq.$\left(  \ref{WDW}\right)  $ by $\Psi^{\ast}\left[  g_{ij}\right]  $ and
functionally integrate over the three spatial metric $g_{ij}$, we can
write\footnote{See also Ref.\cite{CG} for an application of the method to a
$f\left(  R\right)  $ theory.}\cite{Garattini:2004zu, Garattini:2005kx, Garattini:2005ky}
\begin{equation}
\frac{1}{V}\frac{\int\mathcal{D}\left[  g_{ij}\right]  \Psi^{\ast}\left[
g_{ij}\right]  \int_{\Sigma}d^{3}x\hat{\Lambda}_{\Sigma}\Psi\left[
g_{ij}\right]  }{\int\mathcal{D}\left[  g_{ij}\right]  \Psi^{\ast}\left[
g_{ij}\right]  \Psi\left[  g_{ij}\right]  }=\frac{1}{V}\frac{\left\langle
\Psi\left\vert \int_{\Sigma}d^{3}x\hat{\Lambda}_{\Sigma}\right\vert
\Psi\right\rangle }{\left\langle \Psi|\Psi\right\rangle }=-\frac{\Lambda
}{\kappa}, \label{VEV}%
\end{equation}
where we have also integrated over the hypersurface $\Sigma$ and we have
defined
\begin{equation}
V=\int_{\Sigma}d^{3}x\sqrt{g}%
\end{equation}
as the volume of the hypersurface $\Sigma$ with
\begin{equation}
\hat{\Lambda}_{\Sigma}=\left(  2\kappa\right)  G_{ijkl}\pi^{ij}\pi^{kl}%
-\sqrt{g}^{3}R/\left(  2\kappa\right)  . \label{LambdaSigma}%
\end{equation}
In this form, Eq.$\left(  \ref{VEV}\right)  $ can be used to compute ZPE
provided that $\Lambda/\kappa$ be considered as an eigenvalue of $\hat
{\Lambda}_{\Sigma}$, namely the WDW equation is transformed into an
expectation value computation. In Eq.$\left(  \ref{WDW}\right)  $, $G_{ijkl}$
is the super-metric, $\pi^{ij}$ is the super-momentum,$^{3}R$ is the scalar
curvature in three dimensions and $\Lambda$ is the cosmological constant,
while $\kappa=8\pi G$ with $G$ the Newton's constant. Nevertheless, solving
Eq.$\left(  \ref{VEV}\right)  $ is a quite impossible task, therefore we are
oriented to use a variational approach with trial wave functionals. The
related boundary conditions are dictated by the choice of the trial wave
functionals which, in our case\textbf{,} are of the Gaussian type. Different
types of wave functionals correspond to different boundary conditions. The
choice of a Gaussian wave functional is justified by the fact that ZPE should
be described by a good candidate of the \textquotedblleft\textit{vacuum
state}\textquotedblright{}. To fix the ideas, a variant of the line element
$\left(  \ref{line}\right)  $ will be considered
\begin{equation}
ds^{2}=-N^{2}\left(  r\right)  \frac{dt^{2}}{g_{1}^{2}\left(  E/E_{\mathrm{Pl}%
}\right)  }+\frac{dr^{2}}{\left(  1-\frac{b\left(  r\right)  }{r}\right)
g_{2}^{2}\left(  E/E_{\mathrm{Pl}}\right)  }+\frac{r^{2}}{g_{2}^{2}\left(
E/E_{\mathrm{Pl}}\right)  }\left(  d\theta^{2}+\sin^{2}\theta d\phi
^{2}\right)  , \label{dS}%
\end{equation}
where $N$ is the lapse function and $b\left(  r\right)  $ is subject to the
only condition $b\left(  r_{t}\right)  =r_{t}$. For instance, For the
Schwarzschild case, we find $b\left(  r\right)  =2MG=r_{t}$. For the de Sitter
case (dS), ons gets $b\left(  r\right)  =\Lambda_{dS}r^{3}/3$ and for the
Anti-de Sitter (AdS) case one gets $b\left(  r\right)  =-\Lambda_{AdS}r^{3}%
/3$. The graviton contribution of Eq.$\left(  \ref{VEV}\right)  $ is
\begin{equation}
\frac{\Lambda}{8\pi G}=-\frac{1}{3\pi^{2}}\sum_{i=1}^{2}\int_{E^{\ast}%
}^{+\infty}E_{i}g_{1}\left(  E/E_{\mathrm{Pl}}\right)  g_{2}\left(
E/E_{\mathrm{Pl}}\right)  \frac{d}{dE_{i}}\sqrt{\left(  \frac{E_{i}^{2}}%
{g_{2}^{2}\left(  E/E_{\mathrm{Pl}}\right)  }-m_{i}^{2}\left(  r\right)
\right)  ^{3}}dE_{i}, \label{LoverG}%
\end{equation}
where $E^{\ast}$ is the value which annihilates the argument of the root and
where we have defined two r-dependent effective masses $m_{1}^{2}\left(
r\right)  $ and $m_{2}^{2}\left(  r\right)  $
\begin{equation}
\left\{
\begin{array}
[c]{c}%
m_{1}^{2}\left(  r\right)  =\frac{6}{r^{2}}\left(  1-\frac{b\left(  r\right)
}{r}\right)  +\frac{3}{2r^{2}}b^{\prime}\left(  r\right)  -\frac{3}{2r^{3}%
}b\left(  r\right) \\
\\
m_{2}^{2}\left(  r\right)  =\frac{6}{r^{2}}\left(  1-\frac{b\left(  r\right)
}{r}\right)  +\frac{1}{2r^{2}}b^{\prime}\left(  r\right)  +\frac{3}{2r^{3}%
}b\left(  r\right)
\end{array}
\right.  \quad\left(  r\equiv r\left(  x\right)  \right)  . \label{masses}%
\end{equation}
We refer the reader to Refs. \cite{RGGM,RGGM1} for the deduction of these
expressions. It is immediate to recognize that the induced cosmological
constant is no longer a constant but is induced by quantum fluctuations with
the help of Eq.$\left(  \ref{LoverG}\right)  $. Therefore, if we do the
following identification
\begin{equation}
\rho(r)=\frac{\Lambda(r)}{8\pi G} \label{rho}%
\end{equation}
we have the possibility to probe different energy density profiles induced by
quantum fluctuations of the gravitational field itself. To be more explicit,
we choose \cite{RGGM}:
\begin{equation}
g_{1}\left(  E/E_{\mathrm{Pl}}\right)  =(1+\beta\frac{E}{E_{\mathrm{Pl}}%
}+\delta\frac{E^{2}}{E_{\mathrm{Pl}}^{2}}+\gamma\frac{E^{3}}{E_{\mathrm{Pl}%
}^{3}})\exp(-\alpha\frac{E^{2}}{E_{\mathrm{Pl}}^{2}})\qquad g_{2}\left(
E/E_{\mathrm{Pl}}\right)  =1. \label{g1}%
\end{equation}
We can recognize two relevant cases:

\begin{description}
\item[a)] $m_{1}^{2}\left(  r\right)  =-m_{2}^{2}\left(  r\right)  =m_{0}%
^{2}\left(  r\right)  ,$

\item[b)] $m_{1}^{2}\left(  r\right)  =m_{2}^{2}\left(  r\right)  =m_{0}%
^{2}\left(  r\right)  .$
\end{description}

When condition a) is satisfied, this means that we are describing the
Schwarzschild, Schwarzschild-de Sitter and Schwarzschild-Anti de Sitter cases
in proximity of the throat. On the other hand, when condition b) is satisfied,
we are describing the Minkowski, de Sitter and anti-de Sitter cases. For our
purposes, the case b) is the most significant, especially if we fix our
attention to the de Sitter case which, in static coordinates is simply
described by $b\left(  r\right)  =\Lambda_{dS}r^{3}/3$. In this situation the
effective masses of Eq.$\left(  \ref{masses}\right)  $ take the form
\begin{equation}
m_{1}^{2}\left(  r\right)  =m_{2}^{2}\left(  r\right)  =\frac{6}{r^{2}%
}-\Lambda_{dS},\qquad r\in\left(  0,r_{C}\right]
\end{equation}
with $r_{C}=\sqrt{3/\Lambda_{dS}}$. Defining the dimensionless variable
\begin{equation}
x=\frac{L_{P}}{r}\sqrt{6-\Lambda_{dS}r^{2}},
\end{equation}
we can use the following expression
\begin{equation}
\frac{\Lambda}{8\pi GE_{P}^{4}}=C_{1}+C_{2}x^{2}+\left[  C_{3}-\frac{1}%
{8\pi^{2}}\ln\left(  x^{2}\alpha/4\right)  \right]  x^{4}+O(x^{5}),
\label{LambdaEMxpi}%
\end{equation}
which is valid for $x\ll1$. Assuming $r\gg L_{P}$ and $\Lambda r^{2}=O(1)$,
one gets at the leading order
\begin{equation}
\frac{\Lambda}{8\pi GE_{P}^{4}}=C_{1}+C_{2}\left(  \frac{L_{P}}{r}\right)
^{2}\left(  6-\Lambda_{dS}r^{2}\right)  =C_{1}-6C_{2}\Lambda_{dS}L_{P}%
^{2}+6C_{2}\frac{L_{P}^{2}}{r^{2}}, \label{DGInd}%
\end{equation}
where
\begin{equation}
C_{1}=\frac{-8\alpha^{3/2}-6\sqrt{\pi}\alpha\beta-15\sqrt{\pi}\gamma
-16\sqrt{\alpha}\delta}{8\pi^{2}\alpha^{7/2}},
\end{equation}%
\begin{equation}
C_{2}=+\frac{4\alpha^{3/2}+2\sqrt{\pi}\alpha\beta+3\sqrt{\pi}\gamma
+4\sqrt{\alpha}\delta}{8\pi^{2}\alpha^{5/2}}%
\end{equation}
and
\begin{equation}
C_{3}=\frac{-\alpha^{3/2}-2\gamma_{E}\alpha^{3/2}+2\sqrt{\pi}\alpha\beta
+\sqrt{\pi}\gamma+2\sqrt{\alpha}\delta}{16\pi^{2}\alpha^{3/2}}.
\end{equation}
Because of the identification $\left(  \ref{rho}\right)  $, we have obtained a
Dev-Gleiser-like energy density profile. In the next section we will apply
Gravity's Rainbow to the TOV equation.

\bibliographystyle{ieeetr}

\end{document}